# UNIVERSITÉ DE GENÈVE

<u>SCHOLA GENEVENSIS MDLIX</u>

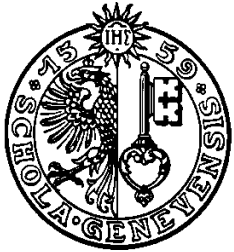

# Ergodic Properties of the Spin – Boson System

*V. Jakšić* [1,2] *and C.-A. Pillet* [3]

[1] Isaac Newton Institute, University of Cambridge, 20 Clarkson Road, Cambridge, CB3 0EH, UK
[2] Institute for Mathematics and its Applications, University of Minnesota, 514 Vincent Hall, 55455–0436 Minneapolis, Minnesota, U.S.A.
[3] Département de Physique Théorique, Université de Genève, CH–1211 Genève 4, Switzerland

**Abstract.** We investigate the dynamics of a 2-level atom (or spin $\frac{1}{2}$) coupled to a mass-less bosonic field at positive temperature. We prove that, at small coupling, the combined quantum system approaches thermal equilibrium. Moreover we establish that this approach is exponentially fast in time. We first reduce the question to a spectral problem for the Liouvillean, a self-adjoint operator naturally associated with the system. To compute this operator, we invoke Tomita-Takesaki theory. Once this is done we use complex deformation techniques to study its spectrum. The corresponding zero temperature model is also reviewed and compared.

## 1. Introduction

In this paper we consider the dissipative dynamics of a quantum mechanical 2-level system — the spin — characterized by its two eigenstates of energy $e_\pm = \pm 1$. More specifically we investigate the long time behavior of the dynamics of a spin $\frac{1}{2}$ allowed to interact with a large reservoir. The reservoir is an infinitely extended gas of free, mass-less bosons at positive temperature without Bose-Einstein condensate. We prove that, for sufficiently small coupling, the interacting *spin-boson* system has strong ergodic properties. In particular it approaches thermal equilibrium exponentially fast. Moreover, the equilibrium state is the unique KMS state of the joint system at the temperature of the heat bath.

The spin-boson system is a simple, yet physically acceptable model for a variety of phenomena related to dissipative quantum tunneling. The literature on the subject is enormous. Let us only mention the review article [LCD] as an excellent introduction to the physical aspects of the model. Also [A1], [A2], [AM], [FNV], [D1], [D2], [HS1], [HS2], [MA], [PU], [SD], [SDLL], [RO1], [RO2] is a non-exhaustive list of related mathematical investigations.

The present work is largely based on results previously obtained by the authors in [JP1]. There we have developed perturbative tools suitable for the study of quantum systems with discrete, possibly infinite, set of energy levels $\{e_i\}$, linearly coupled to a free heat bath at positive temperature. Unfortunately, the general discussion of such systems tends to be technical. For this reason we prefer to restrict ourselves here to a simple model, allowing us to give a more transparent exposition of the underlying ideas. The adaptation of these ideas to more general models will be presented in a subsequent paper [JP2]. For the interested reader, we also present a detailed comparison between our positive temperature model and the more familiar zero temperature spin-boson model. Most of the questions answered in this paper are still open problems at zero temperature.

Our argument splits in three conceptually distinct parts: First we formulate an appropriate generalization of Koopman's Lemma to dynamical systems arising from quantum mechanics. This allows us to reduce ergodic properties of the system to spectral problems for a distinguished self-adjoint operator: The Liouvillean. This operator is defined in abstract terms, and we must invoke Tomita-Takesaki's theory to actually compute it. Once the Liouvillean is known, we apply complex deformation techniques to obtain the relevant spectral informations. On the technical level, one of the main difficulties is the identification of the Liouvillean. The required results of Tomita-Takesaki theory are readily available in the literature. The final step of the proof boils down to an application of the results in [JP1]. An analysis of resonances reveals the basic mechanism of thermal relaxation. In particular, Einstein's A–B law emerges as Fermi's Golden Rule for the resonances of the Liouvillean.

**Acknowledgments.** This paper is part of a program suggested to us by I.M. Sigal, to whom we are grateful. We also thank V. Bach, H. Spohn and P.A. Martin for useful discussions. V.J. is grateful to J.P. Eckmann for hospitality at the University of Geneva, where part of this work was done, and Fonds National Suisse for financial support. At early stage of this work, C.-A.P. was visiting the Institute for Mathematics and its Applications at the University of Minnesota. Its research was also supported by the Fonds National Suisse.

## 2. The model

In this section we define the spin-boson model. We first introduce the isolated spin and the free reservoir, and discuss their thermal equilibrium states. Recall that a *state* of a quantum system is a normalized positive linear functional $\mathcal{S}$ on its algebra of observables. A *vector state* is a state of the form $\mathcal{S}(A) = (\Psi, A\Psi)$, for some unit vector $\Psi$. More generally, a state $\mathcal{S}$ is called *normal* if there exists a density matrix $\varrho$, a positive trace class operator of unit trace, such that $\mathcal{S}(A) = \text{Tr}(\varrho A)$.

The Hilbert space of the isolated spin is $\mathfrak{H}_s \equiv \mathbf{C}^2$. Denoting by $\sigma_x$, $\sigma_y$ and $\sigma_z$ the usual Pauli matrices, we may choose its Hamiltonian to be $h_s \equiv \sigma_z$. The eigenenergies of the spin are $e_\pm = \pm 1$, and we denote the corresponding eigenstates by $\psi_\pm$. Finally, observables of the spin are elements of $M_2$, the algebra of all complex $2 \times 2$ matrices. At inverse temperature $\beta$, the equilibrium state of the spin is the normal state defined by the Gibbs Ansatz

$$\mathcal{S}_s^\beta(A) \equiv \frac{1}{Z_s^\beta} \text{Tr}\left(\exp(-\beta h_s) A\right), \tag{2.1}$$

where $Z_s^\beta$ is a normalization factor. The zero-temperature equilibrium state is obtained in the limit $\beta \uparrow \infty$: It is the vector state corresponding to the ground state of $h_s$

$$\mathcal{S}_s^\infty(A) \equiv (\psi_-, A\psi_-). \tag{2.2}$$

At vanishing density, the Hilbert space of the free reservoir is the symmetric Fock space constructed over $L^2(\mathbf{R}^3)$, which we denote by $\mathfrak{H}_b$ (we work in the momentum representation, thus elements of $L^2(\mathbf{R}^3)$ are always functions of the momentum $\mathbf{k}$ of an individual boson). Since the bosons are non-interacting, the dynamics of the reservoir is completely determined by the energy $\omega(\mathbf{k})$ of a single boson with momentum $\mathbf{k} \in \mathbf{R}^3$. This dynamics is implemented by the strongly continuous unitary group

$$\exp\left(ih_b t\right) \equiv \Gamma\left(\exp(i\omega t)\right),$$

which, by definition, acts on the $N$-particle subspace of $\mathfrak{H}_b$ as the $N$-fold tensor product of the one-boson propagator $\exp(i\omega t)$. In terms of the usual creation and annihilation operators

$a^*(\mathbf{k})$, $a(\mathbf{k})$, the Hamiltonian $h_b$ is given by the familiar formula

$$h_b = \mathrm{d}\Gamma(\omega) \equiv \int_{\mathbf{R}^3} \omega(\mathbf{k})\, a^*(\mathbf{k})\, a(\mathbf{k})\, d^3k.$$

In the sequel we restrict ourselves to the physically important case

$$\omega(\mathbf{k}) \equiv |\mathbf{k}|.$$

However, our method easily accommodates other dispersion laws, as long as the bosons remain mass-less. In this case $h_b$ has a simple eigenvalue 0 corresponding to the Fock vacuum $\Omega_b$, the remaining part of its spectrum is absolutely continuous and fills the positive real axis. The observables of the reservoir are the field operators

$$\phi(f) \equiv \frac{1}{\sqrt{2}} \int_{\mathbf{R}^3} \big(a(\mathbf{k}) + a^*(\mathbf{k})\big)\, f(\mathbf{k})\, d^3k,$$

$$\pi(f) \equiv \frac{1}{i\sqrt{2}} \int_{\mathbf{R}^3} \big(a(\mathbf{k}) - a^*(\mathbf{k})\big)\, f(\mathbf{k})\, d^3k,$$

which satisfy the canonical commutation relations (CCR). A mathematically more convenient set of observables is provided by the Weyl system

$$W(f) \equiv \exp\big(i\varphi(f)\big),$$

where $\varphi(f)$ is the self-adjoint (Segal) field operator defined by

$$\varphi(f) \equiv \frac{1}{\sqrt{2}} \int_{\mathbf{R}^3} \big(a(\mathbf{k})\overline{f(\mathbf{k})} + a^*(\mathbf{k})f(\mathbf{k})\big)\, d^3k,$$

for $f \in L^2(\mathbf{R}^3)$. The operators $W(f)$ are unitary on $\mathfrak{H}_b$, and satisfy a disguised form of CCR: The Weyl relation

$$W(f_1)W(f_2) = \exp\big(-i\,\mathrm{Im}\,(f_1, f_2)/2\big)\, W(f_1 + f_2). \tag{2.3}$$

The dynamics of the reservoir induces a Bogoliubov transformation

$$\exp\big(ih_b t\big)\, W(f)\, \exp\big(-ih_b t\big) = W(\exp(-i\omega t)f),$$

of the Weyl system.

It is a well known fact that thermal equilibrium states of extended systems arise in the thermodynamic limit, starting with a system restricted to a finite box $\Lambda \subset \mathbf{R}^3$. For such a confined system, the grand canonical ensemble yields a well defined state. Equilibrium states of the extended system are constructed as weak-$*$ limits of these states as $\Lambda \uparrow \mathbf{R}^3$. At positive temperature, the equilibrium states obtained in this way have a positive density i.e., an infinite number of particles: They do not fit in the original Fock space (in more technical terms they are not normal). For practical purposes however, it is convenient to restore the familiar Hilbert space framework. This can be achieved by an appropriate choice of the representation of CCR. Let us briefly review some facts about such representations.

Let $\mathfrak{D} \subset L^2(\mathbf{R}^3)$ be a dense subspace. Weyl's algebra over $\mathfrak{D}$ is the C*– algebra generated by the set $\{W(f) : f \in \mathfrak{D}\}$. A representation $(\mathcal{H}, \pi)$ of this algebra is called *regular* if the functions
$$\mathbf{R} \ni \lambda \mapsto \pi\left(W(\lambda f)\right),$$
are strongly continuous for each $f \in \mathfrak{D}$. By Stone's theorem, regularity is equivalent to the existence of a self-adjoint operator $\varphi_\pi(f)$ such that
$$\pi\left(W(\lambda f)\right) = \exp(i\lambda\,\varphi_\pi(f)),$$
for $\lambda \in \mathbf{R}$. One then refers to the $\varphi_\pi(f)$ as the field operators of the representation. A representation is called *cyclic* if for some $\Omega \in \mathcal{H}$ the set $\{\pi\left(W(f)\right)\Omega : f \in \mathfrak{D}\}$ is total in $\mathcal{H}$. To each cyclic representation $(\mathcal{H}, \pi, \Omega)$ of Weyl's algebra, we can associate a *generating functional* defined by
$$\mathfrak{D} \ni f \mapsto (\Omega, \pi\left(W(f)\right)\Omega).$$
Generating functionals of regular, cyclic representations have been characterized by Araki and Segal:

**Theorem 2.1.** *A map $\mathfrak{s}: \mathfrak{D} \to \mathbf{C}$ is the generating functional of a regular, cyclic representation of Weyl's algebra over $\mathfrak{D}$ if and only if the following conditions are satisfied:*

(i) $\mathfrak{s}(0) = 1$.

(ii) *For each $f \in \mathfrak{D}$ the mapping $\mathbf{R} \ni \lambda \mapsto \mathfrak{s}(\lambda f)$ is continuous.*

(iii) *For each finite subset $\{f_1, f_2, \ldots, f_n\} \subset \mathfrak{D}$, and any $z_1, \ldots, z_n \in \mathbf{C}$ one has*
$$\sum_{i,j=1}^n \mathfrak{s}(f_i - f_j) \exp\left(-i\operatorname{Im}(f_i, f_j)/2\right) \bar{z}_i z_j \geq 0.$$

*Furthermore, this representation is unique, up to unitary equivalence.*

At zero chemical potential (which is appropriate for mass-less particles), and in the absence of condensate, the thermodynamic limit leads to the following generating functional for the infinite free Bose gas (see for example [BR2], [LP] or [CA])
$$\mathfrak{s}^\beta(f) \equiv \exp\left(-\frac{1}{4}\int_{\mathbf{R}^3}(1 + 2\rho(\mathbf{k}))\,|f(\mathbf{k})|^2\,d^3k\right), \tag{2.4}$$

for $f$ in the dense subspace $\mathfrak{D}_{loc} \subset L^2(\mathbf{R}^3)$ of functions with compactly supported Fourier transform (i.e., localized in the position representation). In Equation (2.4), the function $\rho(\mathbf{k})$ is the equilibrium momentum distribution of the bosons, and is related to their energy according to Planck's radiation law

$$\rho(\mathbf{k}) = \frac{1}{\exp(\beta\,\omega(\mathbf{k})) - 1}. \tag{2.5}$$

The energy density of the boson gas is strictly positive, and satisfies the Stefan-Boltzmann relation

$$\int \omega(\mathbf{k})\rho(\mathbf{k})\,d^3k \propto \beta^{-4}. \tag{2.6}$$

Let us denote by $\mathcal{A}_{loc}$ the Weyl algebra over $\mathfrak{D}_{loc}$. Since we are dealing with a system at positive density, this is a natural minimal set of observables. By Theorem 2.1, the above functional (2.4) corresponds to a regular, cyclic representation $(\mathcal{H}_b, \pi_{AW}, \Psi_b)$ of $\mathcal{A}_{loc}$. This representation has been explicitly constructed by Araki and Woods (see [AW1], [BR2], [CH] or [LP]), and can described as follows: $\mathcal{H}_b$ is the space of all Hilbert-Schmidt operators on $\mathfrak{H}_b$ equipped with the inner product

$$(X, Y) \equiv \mathrm{Tr}(X^*Y). \tag{2.7}$$

The representant of $W(f)$ acts according to

$$\pi_{AW}(W(f)): X \mapsto W\left((1+\rho)^{1/2} f\right) X W\left(\rho^{1/2} f\right),$$

for any $X \in \mathcal{H}_b$. Finally the cyclic vector is the projection on the Fock vacuum

$$\Psi_b \equiv \Omega_b\,(\Omega_b, \cdot).$$

One easily verifies that the state

$$\mathcal{S}_b^\beta(A) \equiv \left(\Psi_b, \pi_{AW}(A)\,\Psi_b\right),$$

reproduces the functional (2.4), and that the free dynamics has a unitary implementation in the space $\mathcal{H}_b$

$$\pi_{AW}\left(\exp(ih_b t)A\exp(-ih_b t)\right) = \exp(iH_b t)\pi_{AW}(A)\exp(-iH_b t). \tag{2.8}$$

The generator of this group can be explicitly written as

$$H_b: X \mapsto [h_b, X]. \tag{2.9}$$

In the sequel, we will always work in the Araki-Woods representation. Consequently we shall give to the representants $\pi_{AW}(A)$ the status of observables of the boson gas at positive temperature. For reasons which will soon become clear, it is convenient to consider also the von Neumann algebra generated by these representants. Recall that a $C^*$– algebra of operators on a Hilbert space $\mathfrak{H}$ is a von Neumann algebra if it is closed in the weak operator topology. Let $\mathcal{B}$ be a set of bounded operators on $\mathfrak{H}$. We denote by $\mathcal{B}'$ its commutant i.e., the set of bounded operators commuting with all elements of $\mathcal{B}$. If $\mathcal{B}$ is closed under hermitean conjugation, $\mathcal{B}'$ is a von Neumann algebra. Moreover the double commutant $\mathcal{B}''$ is the smallest von Neumann algebra containing $\mathcal{B}$ (see [BR1] or [SA]). We define the algebra of observables of the reservoir at positive temperature by

$$\mathfrak{M}_b \equiv \pi_{AW}\left(\mathcal{A}_{loc}\right)''. \tag{2.10}$$

**Remark 1.** The map $\Phi \otimes \bar{\Psi} \mapsto \Phi(\Psi, \cdot)$ provides an isomorphism between $\mathfrak{H}_b \otimes \mathfrak{H}_b$ and $\mathcal{H}_b$. In the sequel we shall identify this two spaces without further mention. For example, the formulae

$$H_b = h_b \otimes I - I \otimes h_b,$$
$$\pi_{AW}(W(f)) = W\left((1+\rho)^{1/2} f\right) \otimes W\left(\rho^{1/2} \bar{f}\right),$$

directly follow from Equation (2.9) under this identification.

In Equation (2.4), the limit $\beta \uparrow \infty$ yields the generating functional

$$\mathfrak{s}^\infty(f) \equiv \exp\left(-\frac{1}{4}\int_{\mathbf{R}^3} |f(\mathbf{k})|^2 \, d^3k\right) = \left(\Omega_b, W(f)\Omega_b\right),$$

which extends to arbitrary $f \in L^2(\mathbf{R}^3)$. Thus, as expected, we recover the original Fock space representation and, here again, the zero-temperature equilibrium state is the vector state associated with the ground state of the system. In this limiting case the density of the gas vanishes (see (2.5) and (2.6)), and a natural set of observables is the full Weyl algebra $\mathcal{A}$ over $L^2(\mathbf{R}^3)$. Note that this $C^*$– algebra is irreducible i.e., $\mathcal{A}'' = \mathcal{L}(\mathfrak{H}_b)$ the space of all bounded linear operators on $\mathfrak{H}_b$.

We are now ready to define the spin-boson model. At zero temperature, the Hilbert space of the joint system is

$$\mathfrak{H} \equiv \mathfrak{H}_s \otimes \mathfrak{H}_b,$$

and its free Hamiltonian is

$$h_0 \equiv h_s \otimes I + I \otimes h_b.$$

The coupling of the two subsystems is achieved by addition of an interaction term, namely

$$h_\lambda \equiv h_0 + \lambda\, q \otimes \varphi(\alpha), \tag{2.11}$$

where $\lambda$ is a real constant, $q \equiv \sigma_x$, and $\alpha \in L^2(\mathbf{R}^3)$. If

$$(1 + \omega^{-1/2})\alpha \in L^2(\mathbf{R}^3), \tag{2.12}$$

then, by standard estimates (see [GJ1], Section 1.2), the interaction term $q \otimes \varphi(\alpha)$ is infinitesimally small with respect to $h_0$. Thus the operator defined by Equation (2.11) is essentially self-adjoint on $\mathfrak{H}_s \otimes D(h_b)$. For simplicity we will also denote by $h_\lambda$ its self-adjoint extension. The dynamics of the model is given by

$$\tilde{\tau}_\lambda^t \colon A \mapsto \exp(ih_\lambda t)\, A\, \exp(-ih_\lambda t).$$

Generally, the algebra $M_2 \otimes \mathcal{A}$ is not invariant under $\tilde{\tau}_\lambda$. Even worse, $\tilde{\tau}_\lambda$ is not continuous in the natural topology of this algebra. To obtain a decent dynamics we must extend the set of observables to the enveloping von Neumann algebra which, by the last remark of the previous paragraph, is $\mathcal{L}(\mathfrak{H})$. Since the function $t \mapsto \tilde{\tau}_\lambda^t(A)$ is continuous in the weak operator topology, Hypothesis (2.12) ensures that the spin-boson model defines a W*-dynamical system $(\mathcal{L}(\mathfrak{H}), \tilde{\tau}_\lambda)$ for any $\lambda \in \mathbf{R}$. Under the stronger condition

$$(1 + \omega^{-1})\alpha \in L^2(\mathbf{R}^3), \tag{2.13}$$

the spectrum of the spin-boson Hamiltonian (2.11) is given by

$$\sigma(h_\lambda) = [e_-(\lambda), \infty[,$$

and $e_-(\lambda)$ is a simple eigenvalue (see e.g. [SP]). We denote the associated normalized eigenvector by $\Psi_\lambda$. The equilibrium state of the spin-boson system at zero temperature is, by definition, the vector-state defined by

$$\mathcal{S}_\lambda^\infty(A) \equiv \left(\Psi_\lambda, A\Psi_\lambda\right). \tag{2.14}$$

Although very natural, this definition is ultimately justified by the fact that, on $M_2 \otimes \mathcal{A}_{loc}$, the state $\mathcal{S}_\lambda^\infty$ is the weak*-limit as $\beta \uparrow \infty$, of the thermal equilibrium states $\mathcal{S}_\lambda^\beta$ to be defined below (see [SP]).

At positive temperature, the Hilbert space of the joint system is $\mathfrak{H}_s \otimes \mathcal{H}_b$, and its free dynamics is generated by the Hamiltonian

$$H_0 \equiv h_s \otimes I + I \otimes H_b.$$

Denoting by $\varphi_{AW}(f)$ the field operators of the Araki-Woods representation, we define the Hamiltonian of the interacting system by

$$H_\lambda \equiv H_0 + \lambda q \otimes \varphi_{AW}(\alpha). \tag{2.15}$$

From the physical point of view, this is just a rephrasing of Definition (2.11) in a different representation. However, in a more mathematical perspective, the existence of an intertwining relation of the type (2.8) between (2.11) and (2.15) is a difficult question which, in our opinion, requires some information on the thermodynamic limit $\Lambda \uparrow \mathbf{R}^3$ (see for example the discussion in Section 5.2.5 of [BR2] or Section V.1.4 of [HA]). Since this problem is of little physical relevance we will not pay more attention to it and accept (2.15) as a definition of the model at positive temperature.

In [JP1] we proved that $H_\lambda$ is essentially self-adjoint on $\mathfrak{H}_s \otimes D(h_b) \otimes D(h_b)$ for any $\lambda \in \mathbf{R}$, provided
$$(\omega + \omega^{-1})\alpha \in L^2(\mathbf{R}^3).$$
Again we shall use the same symbol to denote its self-adjoint extension. Note that in this case the interaction $q \otimes \varphi_{AW}(\alpha)$ is not $H_0$-bounded. Under the above assumption, it is well known that
$$\tau_\lambda^t \colon A \mapsto \exp(iH_\lambda t)\, A\, \exp(-iH_\lambda t),$$
maps the von Neumann algebra
$$\mathfrak{M} \equiv M_2 \otimes \mathfrak{M}_b = \left(M_2 \otimes \pi_{AW}\left(\mathcal{A}_{loc}\right)\right)'',$$
into itself (see [FNV] or [SP] for example, but also Section 6). Since on the other hand the function $t \mapsto \tau_\lambda^t(A)$ is weakly continuous (in fact it is continuous in the $\sigma$-strong $*$ topology), the spin-boson model at positive temperature also defines a W$^*$- dynamical system $(\mathfrak{M}, \tau_\lambda)$. Thermal equilibrium states of such systems are characterized by the KMS condition.

**Definition 2.2.** *Let $(\mathfrak{N}, \tau)$ be a W$^*$- dynamical system, and $\beta > 0$. A state $\mathcal{S}$ on $\mathfrak{N}$ is a $(\tau, \beta)$-KMS state if it satisfies:*

*(i) $\mathcal{S}$ is normal.*

*(ii) For any $A, B \in \mathfrak{N}$ there exists a function $F_{A,B}(z)$, analytic in the strip*
$$\{z : 0 < \mathrm{Im}(z) < \beta\},$$
*continuous on its closure, and satisfying the KMS boundary conditions*
$$F_{A,B}(t) = \mathcal{S}(A\,\tau^t(B)),$$
$$F_{A,B}(t + i\beta) = \mathcal{S}(\tau^t(B)A),$$
*for $t \in \mathbf{R}$.*

In Section 6, we shall prove

**Proposition 2.3.** *For any $\lambda \in \mathbf{R}$ and $\beta > 0$, there exists a unique $(\tau_\lambda, \beta)$-KMS state $S_\lambda^\beta$ on $\mathfrak{M}$.*

**Remark 2.** In the sequel we refer to $\lambda$ as the *friction constant*, and to $\alpha$ as the *form factor*.

## 3. Results

According to the previous section we shall, from now on, assume that

**(H1)** *The form factor $\alpha$ in Equations (2.11) and (2.15) satisfies*

$$(\omega + \omega^{-1})\alpha \in L^2(\mathbf{R}^3).$$

For technical reasons related to the use of complex deformation techniques in [JP1], we also need a regularity assumption on the form factor $\alpha$. To state this hypothesis we need some additional notation. If $\mathfrak{h}$ is a Hilbert space, we denote by $H^2(\delta, \mathfrak{h})$ the Hardy class of $\mathfrak{h}$-valued functions on the strip

$$\mathfrak{S}(\delta) \equiv \{z : |\mathrm{Im}(z)| < \delta\}.$$

The Hilbert space $H^2(\delta, \mathfrak{h})$ consists of all analytic functions $f \colon \mathfrak{S}(\delta) \to \mathfrak{h}$ satisfying

$$\|f\|_{H^2(\delta, \mathfrak{h})} \equiv \sup_{|a| < \delta} \int_{-\infty}^{\infty} \|f(x + ia)\|_{\mathfrak{h}}^2 \, dx < \infty.$$

Let $\mathrm{S}^2$ denote the unit sphere in $\mathbf{R}^3$. Given a function $f$ on $\mathbf{R}^3$, we define a new function $\widetilde{f}$ on $\mathbf{R} \times \mathrm{S}^2$ by the formula

$$\widetilde{f}(s, \hat{k}) \equiv \begin{cases} -|s|^{1/2} \overline{f}(|s|\hat{k}) & \text{if } s < 0, \\ s^{1/2} f(s\hat{k}) & \text{if } s \geq 0. \end{cases} \tag{3.1}$$

Our central technical hypothesis is:

**(H2)** *There exists $0 < \delta < 2\pi/\beta$ such that*

$$\tilde{\alpha} \in H^2(\delta, L^2(\mathrm{S}^2)).$$

Finally we must assume that the spin effectively couples to the reservoir at Bohr's frequency $\Delta \omega = |e_+ - e_-| = 2$,

**(H3)** $\displaystyle \int_{\mathrm{S}^2} |\alpha(2\hat{k})|^2 \, d\sigma(\hat{k}) > 0$,

where $d\sigma$ is the surface measure on $S^2$. Conditions (H1)–(H3) are satisfied, for example, by the function $\alpha(\mathbf{k}) = \sqrt{|\mathbf{k}|}\exp(-|\mathbf{k}|^2)$. More general conditions will be discussed in [JP2].

We are now ready to formulate the problem of return to equilibrium. Our discussion is motivated by the work of Robinson ([RO1], [RO2]).

**Definition 3.1.** *The spin-boson system at zero temperature has the property of return to equilibrium if*
$$\lim_{|t|\to\infty} \mathcal{S}\left(\tilde{\tau}_\lambda^t(A)\right) = \mathcal{S}_\lambda^\infty(A), \tag{3.2}$$
*for any normal state $\mathcal{S}$ and any $A \in M_2 \otimes \mathcal{A}_{loc}$.*

**Definition 3.2.** *The spin-boson system at positive temperature has the property of return to equilibrium if*
$$\lim_{|t|\to\infty} \mathcal{S}\left(\tau_\lambda^t(A)\right) = \mathcal{S}_\lambda^\beta(A), \tag{3.3}$$
*for any normal state $\mathcal{S}$ and $A \in \mathfrak{M}$.*

We remark that, whenever Relation (3.2) or (3.3) holds, one would also like to know the rate at which the limit is achieved.

In the zero-temperature case, the question of return to equilibrium is still an open and, we believe, an outstanding problem of mathematical physics. We will discuss this problem and related difficulties below. For additional informations we refer the reader to [HE], [SC] and [HS1].

Let us now state the main result of this paper which, in view of the difficulties encountered in the zero-temperature case, comes perhaps as a surprise.

**Theorem 3.3.** *Suppose that Hypotheses (H1)—(H3) hold. Then, for $\beta > 0$, there exists a constant $\Lambda(\beta) > 0$, depending only on the form factor $\alpha$, and such that the spin-boson system has the property of return to equilibrium for any real $\lambda$ satisfying $0 < |\lambda| < \Lambda(\beta)$.*

**Remark 1.** An immediate consequence of this theorem is that, for any density matrix $\rho$ on the space $\mathfrak{H}_s \otimes \mathcal{H}_b$ and any $X \in M_2$, we have
$$\lim_{\lambda\to 0}\lim_{t\to\infty} \operatorname{Tr}\left(\rho\,\tau_\lambda^t(X \otimes I)\right) = \frac{\operatorname{Tr}(\exp(-\beta h_s)\,X)}{\operatorname{Tr}(\exp(-\beta h_s))}.$$
For a related discussion in the framework of master equations, see [D1] and [D3], Equation 5.13.

**Remark 2.** We can recover the zero-temperature model in the limit $\beta \uparrow \infty$. Indeed, the partial trace of any normal state $\mathcal{S}$ of the positive temperature model over the third space in $\mathfrak{H}_s \otimes \mathfrak{H}_b \otimes \mathfrak{H}_b$ is a normal state $\widetilde{\mathcal{S}}$ of the zero temperature model. In fact any such state can be obtained in this way. Moreover one can show that for any normal state $\mathcal{S}$,

$$\lim_{\beta \uparrow \infty} \mathcal{S}_\lambda^\beta \left( X \otimes \pi_{AW}(W(f)) \right) = \mathcal{S}_\lambda^\infty \left( X \otimes W(f) \right),$$

$$\lim_{\beta \uparrow \infty} \mathcal{S} \left( \tau_\lambda^t(X \otimes \pi_{AW}(W(f))) \right) = \widetilde{\mathcal{S}} \left( \widetilde{\tau}_\lambda^t(X \otimes W(f)) \right),$$

hold for arbitrary $X \in M_2$ and $f \in \mathfrak{D}_{loc}$. However this limit is quite singular, and our argument can't avoid the constant $\Lambda(\beta)$ of Theorem 3.3 to vanish as $\beta \uparrow \infty$. Thus, our result does not yield any new information concerning the zero-temperature spin-boson system.

**Theorem 3.4.** *Suppose that Hypotheses (H1)—(H3) hold and let $\Lambda(\beta)$ be as in Theorem 3.3. There exist a norm dense set of normal states $\mathcal{N}_0$ and a strongly dense $*$-algebra $\mathfrak{M}_0 \subset \mathfrak{M}$ so that, for $|\lambda| < \Lambda(\beta), \mathcal{S} \in \mathcal{N}_0$ and $A \in \mathfrak{M}_0$, one has*

$$\left| \mathcal{S}\left(\tau_\lambda^t(A)\right) - \mathcal{S}_\lambda^\beta(A) \right| = O\left(e^{-\gamma(\lambda)|t|}\right), \tag{3.4}$$

*as $|t| \to \infty$. The function $\gamma(\lambda)$ is strictly positive for $0 < |\lambda| < \Lambda(\beta)$, and satisfies*

$$\gamma(\lambda) = \lambda^2 \frac{4\pi}{\text{th}\beta} \int_{S^2} |\alpha(2\hat{k})|^2 \, d\sigma(\hat{k}) + O(\lambda^4),$$

*as $\lambda \to 0$.*

**Remark 3.** $\gamma(\lambda)$ is the negative imaginary part of the complex resonance of the Liouvillean which is closest to the real axis (see Section 6 for details).

**Remark 4.** For any $X \in M_2$, we have $X \otimes I \in \mathfrak{M}_0$.

The proofs will be presented in Sections 6. We now turn to the promised discussion of the mechanisms behind thermal relaxation. We discuss first the zero-temperature model. In this case the relevant physical process is *radiative decay*. The spin "radiates" its energy into the "frozen" gas. As this energy propagates towards infinity, the interacting system dissipates to its lowest energy state: The ground state. We discuss this process in some detail in the sequel.

The spectrum of the uncoupled Hamiltonian $h_0$ is given by

$$\sigma_{ac}(h_0) = [-1, \infty[,$$
$$\sigma_{sc}(h_0) = \emptyset,$$
$$\sigma_{pp}(h_0) = \{-1, +1\},$$

and the eigenvectors associated to the eigenvalues $e_\pm = \pm 1$ are $\psi_\pm \otimes \Omega_b$. As the coupling is "switched on", the eigenvalue $e_- = -1$ moves along the real axis to a new location $e_-(\lambda)$. It remains simple, merely experiencing what is called the *Lamb shift*. The fate of the other eigenvalue is quite different since it is embedded in the continuous spectrum. We expect $e_+ = 1$ to *turn into a resonance* in the following sense: There are $\varepsilon > 0$, $\eta > 0$ and a dense set of vectors $\mathcal{E} \in \mathfrak{H}$ such that, for $|\lambda| < \varepsilon$ and $\Psi \in \mathcal{E}$, the functions

$$R_\Psi(z) \equiv \left(\Psi, (h_\lambda - z)^{-1}\Psi\right),$$

have a meromorphic continuation from the upper half-plane onto the region $\mathcal{O} \equiv \{z : |z - 1| < \eta\}$. The functions $R_\Psi$ should be regular in $\mathcal{O}$ except for a simple pole at $e_+(\lambda)$, which is independent on the choice of $\Psi$.

The above scenario is still a conjecture. Nevertheless, there exists a well developed formal method, going under the name *time-dependent perturbation theory*, which has been used since the 20's to compute the coefficients in the formal Taylor expansion of $e_+(\lambda)$ (see [DI], [HE] and [SC]). The imaginary part of the first non-trivial term in this expansion is generally known as Fermi's Golden Rule. For the model (2.11), this method yields

$$e_+(\lambda) = 1 + \lambda^2 a^{(2)} + O(\lambda^4),$$
$$-\mathrm{Im}\,(a^{(2)}) = \Gamma_+^\infty \equiv 4\pi \int_{S^2} |\alpha(2\hat{k})|^2\, d\sigma(\hat{k}).$$

The quantity $2\lambda^2 \Gamma_+^\infty$ is, in second order perturbation theory, the probability per unit time for the spin to make a transition $\psi_+ \to \psi_-$ while emitting a boson of frequency $\nu = \Delta\omega/2\pi$ into the reservoir. The corresponding spectral line is not infinitely sharp: By the uncertainty relation between time and energy, its width is $\lambda^2 \Gamma_+^\infty$ (see [WW] and [HE] for more details).

If there is a resonance near $e_+ = 1$ in the above sense, and if for some $\varepsilon_0 > 0$ one has $\mathrm{Im}\,(e_+(\lambda)) < 0$ for $0 < |\lambda| < \varepsilon_0$, then the spectrum of $h_\lambda$ in the interval $]1 - \eta, 1 + \eta[$ is purely absolutely continuous as long as $0 < |\lambda| < \varepsilon_0$. In fact we expect more, namely

$$\begin{aligned}\sigma_{ac}(h_\lambda) &= [e_-(\lambda), \infty[, \\ \sigma_{sc}(h_\lambda) &= \emptyset, \\ \sigma_{pp}(h_\lambda) &= \{e_-(\lambda)\},\end{aligned} \qquad (3.5)$$

the eigenvalue $e_-(\lambda)$ being simple. From this and the fact that

$$\mathcal{S}_\lambda^\infty \left(A\,\tilde{\tau}_\lambda^t(B)\right) = \left(A^*\Psi_\lambda, \exp\left(i(h_\lambda - e_-(\lambda))t\right) B\,\Psi_\lambda\right),$$

an elementary property of absolutely continuous spectrum ([RS3], Section 3, Lemma 2), would imply the *mixing property* of the spin-boson system at zero-temperature:

$$\lim_{|t|\to\infty} \mathcal{S}_\lambda^\infty \left(A\,\tilde{\tau}_\lambda^t(B)\right) = \mathcal{S}_\lambda^\infty(A)\mathcal{S}_\lambda^\infty(B), \qquad (3.6)$$

for any observables $A, B \in M_2 \otimes \mathcal{A}_{loc}$. One easily shows that (3.2) implies (3.6), but the opposite is not true: *At zero-temperature, mixing is strictly weaker than return to equilibrium.* We shall see in Section 4 that the situation is different at positive temperature.

In our opinion, the first step towards a proof of return to equilibrium at zero temperature should be a derivation of (3.5), or at least of the mixing property (3.6). However, in view of the previous discussion, we believe that the question of return to equilibrium cannot be, in any natural way, mapped into a spectral problem, and that a new idea is needed.

Important advances in this direction have been recently made in [HS1], [HS2]. In [HS1], Hübner and Spohn develop scattering theory for the model. Motivated by physical arguments they construct an unbounded identification operator

$$J: D \to \mathfrak{H},$$

with domain $D \subset \mathfrak{H}_b$, such that for $\Phi \in D$ the limits

$$\Omega^\pm \Phi = \operatorname*{s-lim}_{t \to \mp\infty} \exp\left(i(h_\lambda - e_-(\lambda))t\right) J \exp\left(-ih_b t\right) \Phi,$$

exist. The wave operators obtained in this way extend to isometries $\Omega^\pm : \mathfrak{H}_b \to \mathfrak{H}$. They enjoy the usual intertwining property. Moreover if the normal state $\mathcal{S}$ is chosen so that the eigenvectors of the associated density matrix belongs to $\operatorname{Ran}(\Omega^\pm)$, then Relation (3.2) holds as $t \to \mp\infty$. Return to equilibrium thus reduces to *asymptotic completeness*:

$$\operatorname{Ran}(\Omega^+) = \operatorname{Ran}(\Omega^-) = \mathfrak{H}. \tag{3.7}$$

This is a stronger property than return to equilibrium. Nevertheless, besides of its own independent interest, scattering theory appears to be a most natural and elegant way to approach this kind of questions. Needles to say, asymptotic completeness is very hard to prove: Similar scattering problems in mathematical physics, for example spin-wave scattering for Heisenberg model or Haag-Ruelle theory of quantum field scattering (see e.g [RS3]), await a resolution since decades. The only model for which Relation (3.7) is known to hold is the exactly solvable model of a harmonic oscillator linearly coupled to a free Bose field [A1].

We would also like to mention the work [HS2], where Mourre theory is adapted to a spin-boson model with *massive* bosons at zero temperature. It would be very interesting to extend these results to the *mass-less* model discussed here.

When this work was finished, we learned that Bach, Frölich and Sigal announced a result which, specialized to the model (2.11), yields (3.5) under some technical assumptions on $\alpha$, e.g. dilation analyticity ([BFS]). In addition, their results in essence justify the resonance picture sketched above.

In the remaining part of this section we briefly discuss the physical processes at positive temperature. To emphasize the physical content of the model, we shall use its *atom-photon* interpretation (see e.g., [CDG]). The operator $H_0$ has the following spectrum:

$$\sigma_{ac}(H_0) = \mathbf{R},$$
$$\sigma_{sc}(H_0) = \emptyset,$$
$$\sigma_{pp}(H_0) = \{-1, +1\}.$$

Unlike in the zero-temperature case, we expect both eigenvalues to turn into resonances as a result of the coupling. This was rigorously established by the authors in [JP1], where Theorem 2.2 translates into the following statement.

**Theorem 3.5.** *Suppose that Hypotheses (H1)—(H3) hold. Then there exists a dense subspace $\mathcal{E} \subset \mathfrak{H}_s \otimes \mathcal{H}_b$ and, for $0 < \eta < \delta$, a constant $\Lambda(\eta) > 0$ such that if $|\lambda| < \Lambda(\eta)$ and $\Phi, \Psi \in \mathcal{E}$, the functions*

$$z \mapsto (\Phi, (H_\lambda - z)^{-1} \Psi), \tag{3.8}$$

*have a meromorphic continuation from the upper half-plane onto the region*

$$\mathcal{O} \equiv \{z : \operatorname{Im}(z) > -\eta\}.$$

*On $\mathcal{O}$, the functions (3.8) are analytic except for two simple poles located at $E_\pm(\lambda)$. The functions $E_\pm(\lambda)$ are analytic for $|\lambda| < \Lambda(\eta)$. Furthermore the first coefficient in the Taylor expansion*

$$E_\pm(\lambda) = \pm 1 + a_\pm^{(2)} \lambda^2 + \ldots,$$

*is given by*

$$\Gamma_\pm^\beta \equiv -\operatorname{Im}(a_\pm^{(2)}) = 2\pi \frac{\exp(\pm\beta)}{|\operatorname{sh}(\beta)|} \int_{\mathbf{S}^2} |\alpha(2\hat{k})|^2 \, d\sigma(\hat{k}),$$

$$\Pi_\pm^\beta \equiv \operatorname{Re}(a_\pm^{(2)}) = \pm \frac{1}{2} \operatorname{PV}\!\!\int_{\mathbf{R} \times \mathbf{S}^2} \frac{\exp(\pm\beta s/2)}{|\operatorname{sh}(\beta s/2)|} \frac{|\tilde{\alpha}(s, \hat{k})|^2}{2 - s} \, ds \, d\sigma(\hat{k}),$$

*where* PV *stands for Cauchy's principal value.*

As an immediate consequence we obtain that, for small non-zero $\lambda$

$$\sigma_{ac}(H_\lambda) = \mathbf{R},$$
$$\sigma_{sc}(H_\lambda) = \emptyset,$$
$$\sigma_{pp}(H_\lambda) = \emptyset.$$

Moreover we remark that the imaginary part of $E_\pm(\lambda)$ is related to the radiative lifetime of the corresponding atomic state $\psi_\pm$ by Cauchy integral formula (see [JP1] for details). In second order perturbation theory, only processes in which a single photon is either emitted or absorbed are taken into account. In this framework, the coefficients $\Gamma_\pm^\beta$ have a simple physical interpretation: $2\lambda^2 \Gamma_+^\beta$ is the probability per unit time that the two-level atom will emit a photon of frequency $\nu = \Delta\omega/2\pi = 1/\pi$ and make a transition $\psi_+ \to \psi_-$. Similarly $2\lambda^2 \Gamma_-^\beta$ is the probability per unit unit time that the atom will absorb a photon of frequency $\nu$ and make the reverse transition $\psi_- \to \psi_+$. Again, the emission and absorption lines have a finite width given by $\lambda^2 \Gamma_\pm^\beta$. Note that

$$\lim_{\beta \uparrow \infty} \Gamma_-^\beta = 0,$$
$$\lim_{\beta \uparrow \infty} \Gamma_+^\beta = \Gamma_+^\infty,$$

as expected: At zero temperature only the emission process must be taken into account.

Let now $p_\pm$ be the probability for the atom to be in the state $\psi_\pm$, then we must have

$$p_- + p_+ = 1. \tag{3.9}$$

If the entire system is in thermal equilibrium, detailed balance further requires

$$p_+ \Gamma_+^\beta = p_- \Gamma_-^\beta. \tag{3.10}$$

The only solution of the system (3.9)—(3.10) is the canonical Gibbs distribution associated to the Hamiltonian $h_s$ i.e.,

$$p_\pm = \frac{\exp(\pm\beta)}{\exp(\beta) + \exp(-\beta)},$$

as expected. The general form of the coefficients $\Gamma_\pm^\beta$ has been postulated by Einstein ([E], see also [PA] and [P]) in 1917, and is known as Einstein's A–B law. To calculate these coefficients, Dirac ([DI]) developed the aforementioned time-dependent perturbation theory. The notion of resonances discussed above emerged in the early seventies ([AC], [BC], [SI1], [HO], [RS4]) in the context of $N$-body non-relativistic quantum mechanics, as an attempt to find a mathematically satisfactory foundation for time-dependent perturbation theory.

The spectral analysis of $H_\lambda$ is only a first step towards the understanding of the long time behavior of the spin-boson model, and the proof of (3.3) requires a new ingredient. Fortunately we can avoid scattering theory by exploiting the rich algebraic structure associated with the positive temperature model. In the next section we show how this structure can be used to transform the question into a spectral problem.

## 4. Quantum Koopmanism

In classical mechanics, the spectral characterization of dynamics is based on Koopman's lemma: Let $(M, \mu, \varphi)$ be a *classical dynamical system* i.e., $M$ a measure space, $\mu$ a probability measure on $M$ and $\mathbf{R} \ni t \mapsto \varphi^t$ a measurable group of measure preserving transformations of $M$. Assume also that the Hilbert space $L^2(M, d\mu)$ is separable. Then

$$L^2(M, d\mu) \ni f \mapsto f \circ \varphi^t, \tag{4.1}$$

defines a strongly continuous unitary group. Thus a classical dynamical system has an associated self-adjoint operator: The generator $L$ of the group (4.1). It turns out that the spectrum of Koopman's operator $L$ carries important informations on the behavior of the dynamical system (see [CFS] for more details). Note that if $\varphi$ is the Hamiltonian flow associated with the Hamilton function $h$, and if we denote the Poisson bracket by $\{\cdot, \cdot\}$, then $L$ is just the usual Liouville operator: $f \mapsto i\{h, f\}$.

Our aim is to extend this framework to quantum mechanics. Let $\mathfrak{M}$ be a von Neumann algebra. Recall that a state $\mathcal{S}$ on $\mathfrak{M}$ is faithful if $\mathcal{S}(A^*A) = 0$ implies $A = 0$. We shall call $(\mathfrak{M}, \mathcal{S}, \tau)$ a *quantum dynamical system* if $\mathbf{R} \ni t \mapsto \tau^t$ is a weakly continuous group of automorphisms of $\mathfrak{M}$, and $\mathcal{S}$ a faithful normal $\tau$-invariant state. We further denote by $(\mathcal{H}, \pi, \Omega)$ the canonical cyclic representation of $\mathfrak{M}$ (see [BR1], Section 2.3.3) associated to $\mathcal{S}$. The two conditions

$$\begin{aligned} \pi(\tau^t(A)) &= \exp(iLt)\,\pi(A)\,\exp(-iLt), \\ L\Omega &= 0, \end{aligned} \tag{4.2}$$

uniquely determine a self-adjoint operator $L$ on the Hilbert space $\mathcal{H}$. We call $L$ the *Liouvillean* of the system since, as we shall see below, it reduces to the ordinary Liouville operator in the classical Hamiltonian case (see [RO1], where $L$ is called equilibrium Hamiltonian). Note that the second condition in (4.2) is crucial: Many operators satisfy the first condition. In fact if $L_0$ is such an operator, so is $L_0 + V$ for any self-adjoint $V \in \pi(\mathfrak{M})'$.

As we shall see shortly, the Liouvillean is the appropriate generalization of Koopman's operator. To formulate our result we need some definitions.

**Definition 4.1.** *Let $(\mathfrak{M}, \mathcal{S}, \tau)$ be a quantum dynamical system, and denote by $\mathcal{N}$ the set of normal states on $\mathfrak{M}$.*

(i) *$(\mathfrak{M}, \mathcal{S}, \tau)$ is ergodic if, for any $A \in \mathfrak{M}$ and $\mathcal{S}' \in \mathcal{N}$, one has*

$$\lim_{T \to \infty} \frac{1}{2T} \int_{-T}^{T} \mathcal{S}'(\tau^t(A))\, dt = \mathcal{S}(A).$$

(ii) *It is weakly mixing if, for any $A, B \in \mathfrak{M}$, one has*

$$\lim_{T \to \infty} \frac{1}{2T} \int_{-T}^{T} \left| \mathcal{S}(\tau^t(A)B) - \mathcal{S}(A)\,\mathcal{S}(B) \right|^2 dt = 0.$$

(iii) *It is mixing if, for any $A, B \in \mathfrak{M}$, one has*

$$\lim_{t \to \infty} \mathcal{S}(\tau^t(A)B) = \mathcal{S}(A)\,\mathcal{S}(B).$$

(iv) *It returns to equilibrium if, for any $A \in \mathfrak{M}$ and $\mathcal{S}' \in \mathcal{N}$, one has*

$$\lim_{t \to \infty} \mathcal{S}'(\tau^t(A)) = \mathcal{S}(A).$$

At this point, it is instructive to reconsider a classical dynamical system $(M, \mu, \varphi)$. One easily shows that $\mathfrak{M} = L^\infty(M, d\mu)$, $\mathcal{S}(f) = \int f\, d\mu$ and $\tau^t(f) = f \circ \varphi^t$ define a (commutative) quantum dynamical system for which Definition 4.1 reduces to the familiar one ([CFS]). Moreover the corresponding cyclic representation can be constructed in the following way: $\mathcal{H} = L^2(M, d\mu)$, $\pi(f) = f$ (as multiplication operator) and $\Omega = 1$. The Liouvillean $L$ is identical to the original Koopman operator of $(M, \mu, \varphi)$. This motivates the following result.

**Theorem 4.2.** *Let $(\mathfrak{M}, \mathcal{S}, \tau)$ be a quantum dynamical system, $(\mathcal{H}, \pi, \Omega)$ its cyclic representation and $L$ its Liouvillean. Denote also by $P_\Omega$ the orthogonal projection of $\mathcal{H}$ along the cyclic vector $\Omega$. Then,*

(i) *$(\mathfrak{M}, \mathcal{S}, \tau)$ is ergodic if and only if $0$ is a simple eigenvalue of $L$.*

(ii) *It is weakly mixing if and only if $L$ has purely continuous spectrum, except for the simple eigenvalue $0$.*

(iii) *It is mixing if and only if*
$$\text{w} - \lim_{t \to \infty} \exp(-iLt) = P_\Omega. \tag{4.3}$$

(iv) *It returns to equilibrium if and only if it is mixing.*

**Proof.** We start with some basic facts (see [BR1], Sections 2.3.1, 2.4.4 and 2.5.1). Let $\mathfrak{N}$ be a von Neumann algebra on a Hilbert space $\mathfrak{H}$. A vector $\Psi \in \mathfrak{H}$ is called cyclic for $\mathfrak{N}$ if $\mathfrak{N}\Psi$ is dense in $\mathfrak{H}$. It is called separating for $\mathfrak{N}$ if $A \in \mathfrak{N}$ and $A\Psi = 0$ implies $A = 0$.

A vector is separating for $\mathfrak{N}$ if and only if it is cyclic for $\mathfrak{N}'$. A representation $\eta$ of $\mathfrak{N}$ is called faithful if it is a $*$-isomorphism between $\mathfrak{N}$ and $\eta(\mathfrak{N})$. A representation $\eta$ is faithful if and only if $\eta(A) = 0$ implies $A = 0$. A representation is faithful if and only if it satisfies $\|\eta(A)\| = \|A\|$ for any $A \in \mathfrak{N}$. Finally if $\eta$ is a faithful representation of $\mathfrak{N}$ in the Hilbert space $\mathcal{H}$, then a state $\mathcal{S}$ on $\mathfrak{N}$ is normal if and only if there exists a density matrix $\rho$ on $\mathcal{H}$ such that $\mathcal{S}(A) = \mathrm{Tr}(\rho\,\eta(A))$.

The proof of Theorem 4.2 is based on the following argument: The cyclic representation $\pi$ inherits the faithfulness of $\mathcal{S}$, therefore $\Omega$ is not only cyclic but also separating for $\pi(\mathfrak{M})$. It follows that both $\pi(\mathfrak{M})\Omega$ and $\pi(\mathfrak{M})'\Omega$ are dense in $\mathcal{H}$. Let us denote by $\mathcal{N}_0$ the set of vector states arising from vectors in $\pi(\mathfrak{M})'\Omega$. The set of finite convex linear combinations of elements of $\mathcal{N}_0$ is norm dense in $\mathcal{N}$, thus we can replace $\mathcal{N}$ by $\mathcal{N}_0$ in Definition 4.1. Now, for $\mathcal{S}' \in \mathcal{N}_0$, there exists an operator $C \in \pi(\mathfrak{M})'$ such that

$$\begin{aligned}
\mathcal{S}'(\tau^t(A)) - \mathcal{S}(A) &= (C\Omega, \pi(\tau^t(A))C\Omega) - (\Omega, \pi(A)\Omega) \\
&= (\Omega, \pi(\tau^t(A))C^*C\Omega) - (\Omega, \pi(A)\Omega)(\Omega, C^*C\Omega) \\
&= (\pi(A^*)\Omega, \exp(-iLt)(I - P_\Omega)C^*C\Omega),
\end{aligned} \quad (4.4)$$

for any $A \in \mathfrak{M}$. In the same way we can write

$$\begin{aligned}
\mathcal{S}(\tau^t(A)B) - \mathcal{S}(A)\mathcal{S}(B) &= (\Omega, \pi(\tau^t(A)B)\Omega) - (\Omega, \pi(A)\Omega)(\Omega, \pi(B)\Omega) \\
&= (\pi(A^*)\Omega, \exp(-iLt)(I - P_\Omega)\pi(B)\Omega),
\end{aligned} \quad (4.5)$$

for any $A, B \in \mathfrak{M}$. Von Neumann's ergodic theorem (see [CFS] for example) applied to (4.4) and the density of $\pi(\mathfrak{M})\Omega$ and $\pi(\mathfrak{M})'\Omega$ yield a proof of (i). In a completely similar way, RAGE theorem (see [RS3], Theorem XI.115) applied to (4.5) and the density of $\pi(\mathfrak{M})\Omega$ prove (ii). By (4.5), assertion (iii) is an immediate consequence of the density of $\pi(\mathfrak{M})\Omega$. Finally, using (iii), we reduce the proof of assertion (iv) to the equivalence of return to equilibrium with (4.3). This follows directly from (4.4), the density of $\pi(\mathfrak{M})\Omega$ and $\pi(\mathfrak{M})'\Omega$, and a the fact that an arbitrary $P \in \pi(\mathfrak{M})'$ is a linear combination of positive operators.

∎

**Corollary 4.3.** *If the Liouvillean $L$ of a quantum dynamical system has purely absolutely continuous spectrum, except for the simple eigenvalue 0, then this system is mixing.*

**Proof.** This follows from assertion (iii) of Theorem 4.2 and an application of Riemann-Lebesgue's Lemma to the spectral measures of $L$ (see [RS3], Section XI.3, Lemma 2).

∎

**Remark 1.** Let $(\mathfrak{M}, \tau)$ be a W*- dynamical system. Any $(\tau, \beta)$-KMS state $\mathcal{S}^\beta$ is faithful, normal and $\tau$-invariant (see [BR2], Section 5.3.1). It follows that $(\mathfrak{M}, \mathcal{S}^\beta, \tau)$ is a quantum dynamical system. This contrasts with the zero temperature case: At zero temperature an equilibrium state (ground state) generally fails to be faithful. The loss of faithfulness in the limit $\beta \uparrow \infty$ is the source of one of the previously mentioned difficulties in the zero temperature spin-boson model: There is no Koopman Lemma at zero temperature, hence no spectral characterization of the dynamics. This is not a quantum phenomenon, the problem already exists at the classical level.

## 5. Modular Theory

In this section we restrict ourselves to quantum dynamical systems of the form $(\mathfrak{M}, \mathcal{S}^\beta, \tau)$, where $\mathcal{S}^\beta$ is a $(\tau, \beta)$-KMS state. We show how Tomita-Takesaki's theory relates the Liouvillean of the system to its *modular structure*, and how this fact naturally leads to multiplicative and additive perturbation theory of the Liouvillean. We start with a brief review of the basic construction leading to modular theory. For a more detailed exposition from the standpoint of mathematical physics we refer the reader to [AR1] and [BR1],[BR2].

Let $\mathfrak{N}$ be a von Neumann algebra on a Hilbert space $\mathfrak{H}$, and $\Psi \in \mathfrak{H}$ a separating cyclic vector. The formulae
$$\begin{aligned} SA\Psi &= A^*\Psi \quad \text{for} \quad A \in \mathfrak{N}, \\ FB\Psi &= B^*\Psi \quad \text{for} \quad B \in \mathfrak{N}', \end{aligned} \tag{5.1}$$
define two anti-linear operators $S$ and $F$ with dense domains $\mathfrak{N}\Psi$ and $\mathfrak{N}'\Psi$ respectively. Since one easily sees that
$$S \subset F^*, \qquad F \subset S^*, \tag{5.2}$$
the two operators $S$ and $F$ are closable, and we denote their closure with the same symbol. Let us write the polar decomposition of $S$ as
$$S = J\Delta^{1/2}, \tag{5.3}$$
where $J$ is anti-unitary and $\Delta$ self-adjoint and positive. It turns out that $J$ is is an involution: $J^2 = J$. It is the *modular conjugation* of the pair $(\mathfrak{N}, \Psi)$, while $\Delta$ is its *modular operator*. The fundamental theorem of Tomita and Takesaki states that
$$J\Psi = \Psi, \qquad J\mathfrak{N}J = \mathfrak{N}', \tag{5.4}$$
and
$$\Delta^{it}\Psi = \Psi, \qquad \Delta^{it}\mathfrak{N}\Delta^{-it} = \mathfrak{N}, \tag{5.5}$$
for all $t \in \mathbf{R}$.

To make the connection with the quantum dynamical system $(\mathfrak{M}, \mathcal{S}^\beta, \tau)$ note that, according to the previous section, the canonical cyclic representation $(\mathcal{H}, \pi, \Omega)$ associated with $\mathcal{S}^\beta$ is faithful. Thus $\Omega$ is a separating cyclic vector for $\pi(\mathfrak{M})$, and we can apply the above construction to the pair $(\pi(\mathfrak{M}), \Omega)$. By a slight abuse of language, we shall say that the operators $J$ and $\Delta$ obtained in this way are the modular conjugation and modular operator of the system $(\mathfrak{M}, \mathcal{S}^\beta, \tau)$. It follows from Tomita-Takesaki's theorem that the formula

$$\pi^\sharp(A) \equiv J\,\pi(A)\,J,$$

defines an anti-linear representation of $\mathfrak{M}$ on the commutant $\pi(\mathfrak{M})'$. We shall see that this *dual representation* ([AW], [HHW]) plays a fundamental role in our problem. Another deep consequence of Tomita-Takesaki's theorem is that

$$\chi^t(A) \equiv \pi^{-1}\left(\Delta^{it}\,\pi(A)\,\Delta^{-it}\right),$$

defines a group of automorphisms of $\mathfrak{M}$. Takesaki's theorem further asserts that $\chi$ is the unique $\sigma$-weakly continuous group of automorphisms of $\mathfrak{M}$ admitting $\mathcal{S}^\beta$ as a KMS state at inverse temperature $-1$. One easily conclude that

$$\chi^t = \tau^{-\beta t},$$

which we summarize in the next result.

**Proposition 5.1.** *Let $(\mathfrak{M}, \mathcal{S}^\beta, \tau)$ be a quantum dynamical system, and $(\mathcal{H}, \pi, \Omega)$ its cyclic representation. Assume that $\mathcal{S}^\beta$ is a $(\tau, \beta)$-KMS state. Then the Liouvillean $L$ of the system is related to its modular operator $\Delta$ by the formula*

$$\Delta = \exp(-\beta L).$$

The modular structure enjoys very simple covariance properties under unitary transformation (inner automorphisms). In particular the spectrum of the Liouvillean is invariant under such transformation.

**Lemma 5.2.** *Let $(\mathfrak{M}, \mathcal{S}, \tau)$ be a quantum dynamical system, $(\mathcal{H}, \pi, \Omega)$ its cyclic representation, $L$ its Liouvillean and $J$ its modular conjugation. Furthermore let $V$ be a unitary element of $\mathfrak{M}$, and denote by $\gamma$ the associated inner automorphism of $\mathfrak{M}$, i.e.,*

$$\gamma(A) = V^*AV,$$

*for any $A \in \mathfrak{M}$. To $V$ we associate the unitary operator*

$$U \equiv \pi(V)\,\pi^\sharp(V),$$

on $\mathcal{H}$. Then $(\mathfrak{M}, \mathcal{S} \circ \gamma, \gamma^{-1} \circ \tau \circ \gamma)$ is a quantum dynamical system.

(i) Its cyclic representation is given by $(\mathcal{H}, \pi, U\Omega)$.

(ii) Its Liouvillean is $ULU^*$.

(iii) Its modular conjugation is again $J$.

The proof of this Lemma is a simple application of Tomita-Takesaki's theorem, and will be omitted. We end this section with a powerful result which describes how the modular structure is altered by a small perturbation of the dynamics. It is an immediate rephrasing of a well known theorem of Araki [AR2] (see also [BR2], Theorem 5.4.4 and the remarks after it).

**Theorem 5.3.** *Let $(\mathfrak{M}, \mathcal{S}^\beta, \tau)$ be a quantum dynamical system, $(\mathcal{H}, \pi, \Omega)$ its cyclic representation, $L$ its Liouvillean, and $J$ its modular conjugation. Assume that $\mathcal{S}^\beta$ is the unique $(\tau, \beta)$-KMS state of $\mathfrak{M}$. Then, for any self-adjoint $V \in \mathfrak{M}$, the formula*

$$\tau_V^t(A) \equiv \pi^{-1}\left(\exp(i(L+\pi(V))t)\,\pi(A)\,\exp(-i(L+\pi(V))t)\right),$$

*defines a W\*- dynamical system on $\mathfrak{M}$. Furthermore,*

(i) *$\Omega \in D(\exp(-\beta(L+\pi(V))/2))$, and the vector state $\mathcal{S}_V^\beta$ associated with*

$$\Omega_V \equiv \frac{\exp(-\beta(L+\pi(V))/2)\Omega}{\|\exp(-\beta(L+\pi(V))/2)\Omega\|},$$

*is the unique $(\tau_V, \beta)$-KMS state of $\mathfrak{M}$.*

(ii) *The cyclic representation of the perturbed system $(\mathfrak{M}, \mathcal{S}_V^\beta, \tau_V)$ is $(\mathcal{H}, \pi, \Omega_V)$.*

(iii) *Its Liouvillean is $L_V = L + \pi(V) - \pi^\sharp(V)$.*

(iv) *Its modular conjugation is again $J$.*

We are now fully equipped to proceed with the proof of our main result.

## 6. Proofs of Theorem 3.3 and Theorem 3.4

As a warm up, let us describe in details the modular structure of the isolated spin at inverse temperature $\beta$. Recall that the observable algebra is $\mathfrak{M}_s \equiv M_2$. Since it is a factor (i.e., its center $\mathfrak{M}_s \cap \mathfrak{M}_s'$ is trivial), the Gibbs state $\mathcal{S}_s^\beta$ given in Equation (2.1) is the unique KMS state of the system (see [BR2], Theorem 5.3.29). We denote by $(\mathcal{H}_s, \pi_s, \Psi_s)$ the corresponding canonical cyclic representation of $\mathfrak{M}_s$ which, according to the GNS construction, can be realized in the following way: The Hilbert space is

$$\mathcal{H}_s \equiv M_2,$$

with the familiar inner product

$$(X, Y) \equiv \mathrm{Tr}(X^* Y). \tag{6.1}$$

The representant $\pi_s(A)$ acts by left multiplication i.e., for any $A \in \mathfrak{M}_s$ and $X \in \mathcal{H}_s$,

$$\pi_s(A)\colon X \mapsto AX.$$

The unit vector

$$\Psi_s \equiv \frac{1}{\sqrt{Z_s^\beta}} \exp(-\beta h_s/2),$$

is obviously cyclic and separating for $\pi_s(\mathfrak{M}_s)$, and satisfies

$$\mathcal{S}_s^\beta(A) = (\Psi_s, \pi_s(A) \Psi_s).$$

Thus there is a unique self-adjoint operator $L_s$ on $\mathcal{H}_s$ such that

$$\pi_s\left(\exp(ith_s) A \exp(-ith_s)\right) = \exp(itL_s) \pi_s(A) \exp(-itL_s),$$

for $A \in \mathfrak{M}_s$, and

$$L_s \Psi_s = 0.$$

One easily checks that the operator defined by

$$L_s\colon X \mapsto [h_s, X],$$

has the required properties, and therefore is the Liouvillean of the system. It follows from Proposition 5.1 that the modular operator of the pair $(\pi_s(\mathfrak{M}_s), \Omega_s)$ is given by

$$\Delta_s \equiv \exp(-\beta L_s)\colon X \mapsto \exp(-\beta h_s) X \exp(\beta h_s).$$

Going back to the definitions (5.1), (5.3), its modular conjugation is immediately identified as

$$J_s \colon X \mapsto X^*,$$

from which we conclude that

$$\pi_s^\sharp(A) \colon X \mapsto X A^*.$$

Along the same line we shall now describe the modular structure of the isolated boson gas at inverse temperature $\beta$. Recall that the algebra of observables is $\mathfrak{M}_b \equiv \pi_{AW}(\mathcal{A}_{loc})''$, where $\pi_{AW}$ is the Araki-Woods representation corresponding to the equilibrium state $\mathcal{S}_b^\beta$. By construction, the cyclic representation of the reservoir at thermal equilibrium is $(\mathcal{H}_b, \pi_b, \Psi_b)$, where

$$\pi_b(A) = A.$$

As in the case of the isolated spin, it is straightforward to identify the Liouvillean as the operator on $\mathcal{H}_b$ defined by

$$L_b \colon X \mapsto [h_b, X].$$

Note that, in this case, the Liouvillean is identical with the original Hamiltonian $H_b$. Proceeding as before we can write down the modular operator of the pair $(\pi_b(\mathfrak{M}_b), \Psi_b)$,

$$\Delta_b \equiv \exp(-\beta L_b) \colon X \mapsto \exp(-\beta h_b)\, X\, \exp(\beta h_b).$$

Using Definitions (5.1) and (5.3), we see that the modular conjugation $J_b$ is characterized by

$$J_b \exp(-\beta h_b/2)\, W\left(\sqrt{1+\rho}\, f\right) |\Omega\rangle\langle\Omega| W\left(\sqrt{\rho}\, f\right) \exp(\beta h_b/2) = $$
$$W\left(\sqrt{1+\rho}\, f\right)^* |\Omega\rangle\langle\Omega| W\left(\sqrt{\rho}\, f\right)^*.$$

Since $\exp(-\beta h_b/2) = \Gamma(\exp(-\beta\omega/2))$, a well known property of second quantized contractions (see [SI2], Section I.4) gives

$$\exp(-\beta h_b/2)\, W(f)\,\Omega = \exp\left(-\frac{1}{4}\left(f, \left(1 - \mathrm{e}^{-\beta\omega}\right) f\right)\right) W\left(\exp(-\beta\omega/2)f\right)\Omega. \quad (6.2)$$

If $f \in D(\exp(\beta\omega/2))$, it follows from this formula that $W(f)\Omega \in D(\exp(\beta h_b/2))$, and

$$\exp(\beta h_b/2)\, W(f)\,\Omega = \exp\left(\frac{1}{4}\left(f, \left(\mathrm{e}^{\beta\omega} - 1\right) f\right)\right) W\left(\exp(\beta\omega/2)f\right)\Omega. \quad (6.3)$$

Inserting (6.2) and (6.3) in the above characterization of $J_b$ leads to

$$J_b\, W\left(\sqrt{\rho}\, f\right) |\Omega\rangle\langle\Omega| W\left(\sqrt{1+\rho}\, f\right) = W\left(\sqrt{1+\rho}\, f\right)^* |\Omega\rangle\langle\Omega| W\left(\sqrt{\rho}\, f\right)^*,$$

from which it becomes apparent that

$$J_b: X \mapsto X^*.$$

The dual representation is given by

$$\pi_b^\sharp(A): X \mapsto (A X^*)^*,$$

in particular

$$\pi_b^\sharp(\pi_{AW}(W(f))): X \mapsto W\left(\rho^{1/2} f\right)^* X W\left((1+\rho)^{1/2} f\right)^*.$$

From this, a simple calculation shows that

$$\pi_b(\pi_{AW}(W(\sqrt{1+\rho}\,f)))\,\pi_b^\sharp(\pi_{AW}(W(\sqrt{\rho}\,f))^*): X \mapsto W(f)X,$$
$$\pi_b(\pi_{AW}(W(\sqrt{\rho}\,f))^*)\,\pi_b^\sharp(\pi_{AW}(W(\sqrt{1+\rho}\,f))): X \mapsto XW(f),$$

and since $\mathfrak{M}_b' = \pi_b^\sharp(\mathfrak{M}_b)$, the irreducibility of the Fock representation allows us to conclude that $\mathfrak{M}_b \vee \mathfrak{M}_b' = \mathcal{L}(\mathcal{H}_b)$. Hence we recover the well known facts that $\mathfrak{M}_b$ is a factor and that $\mathcal{S}_b^\beta$ is the the unique KMS state of the isolated boson gas at inverse temperature $\beta$.

We can now describe the modular structure of the combined spin-boson system. In the absence of interaction i.e., when $\lambda = 0$, the state

$$\mathcal{S}_0^\beta \equiv \mathcal{S}_s^\beta \otimes \mathcal{S}_b^\beta,$$

is the unique $(\tau_0, \beta)$-KMS state on $\mathfrak{M} = \mathfrak{M}_s \otimes \mathfrak{M}_b$. The quantum dynamical system $(\mathfrak{M}, \mathcal{S}_0^\beta, \tau_0)$ has a canonical cyclic representation $(\mathcal{H}, \pi, \Psi)$ defined by

$$\mathcal{H} = \mathcal{H}_s \otimes \mathcal{H}_b,$$
$$\pi = \pi_s \otimes \pi_b,$$
$$\pi^\sharp = \pi_s^\sharp \otimes \pi_b^\sharp,$$
$$\Psi = \Psi_s \otimes \Psi_b.$$

The vector $\Psi$ is cyclic and separating for $\pi(\mathfrak{M})$, and the corresponding modular operator and modular conjugation are given by

$$\Delta = \Delta_s \otimes \Delta_b,$$
$$J = J_s \otimes J_b.$$

Finally the Liouvillean is

$$L_0 = L_s \otimes I + I \otimes L_b.$$

To obtain the modular structure of the coupled system, we would like to follow the perturbative approach of Theorem 5.3. This is not directly possible, due to the unboundedness of the coupling term $q \otimes \varphi_{AW}(\alpha)$. However, the following twist avoids this complication: Define
$$V_\lambda \equiv \exp\left(i\lambda\, q \otimes \varphi_{AW}\left(i\alpha/\omega\right)\right).$$

One easily checks that Hypotheses (H1)-(H2) imply $\alpha/\omega \in D(\omega^{-1/2})$, from which we can conclude that $V_\lambda \in \mathfrak{M}$. Let us denote by $\gamma_\lambda$ the corresponding inner automorphism of $\mathfrak{M}$. Using Weyl's relation (2.3), an explicit calculation shows that
$$\tau_\lambda^t = \gamma_\lambda^{-1} \circ \xi_\lambda^t \circ \gamma_\lambda,$$

where
$$\xi_\lambda^t(A) = \exp(i(H_0 + T_\lambda)t)\, A\, \exp(-i(H_0 + T_\lambda)t),$$

and $T_\lambda$ is the self-adjoint element of $\mathfrak{M}$ given by
$$T_\lambda = \gamma_\lambda\,(\sigma_z \otimes I) - \sigma_z \otimes I.$$

Since $\xi_0 = \tau_0$, we know from the previous paragraphs that $(\mathfrak{M}, \mathcal{S}_0^\beta, \xi_0)$ is a quantum dynamical system, and that $\mathcal{S}_0^\beta$ is the unique $(\xi_0, \beta)$-KMS state. One further checks that
$$\pi(\xi_\lambda^t(A)) = \exp(i(L_0 + \pi(T_\lambda))t)\,\pi(A)\,\exp(-i(L_0 + \pi(T_\lambda))t),$$

implements the dynamics in the cyclic representation. At this point we are ready to apply Theorem 5.3, which shows that $\xi_\lambda$ defines a quantum dynamical system with unique KMS state a inverse temperature $\beta$. The Liouvillean of this system is given by
$$\widetilde{L}_\lambda \equiv L_0 + \pi(T_\lambda) - \pi^\sharp(T_\lambda).$$

Applying now Lemma 5.2, we conclude that $\tau_\lambda$ also has a unique KMS state $\mathcal{S}_\lambda^\beta$, and that the Liouvillean of $(\mathfrak{M}, \mathcal{S}_\lambda^\beta, \tau_\lambda)$ is given by
$$L_\lambda = \pi(V)\pi^\sharp(V)\widetilde{L}_\lambda \pi(V^*)\pi^\sharp(V^*).$$

This formula can be explicitly evaluated to obtain
$$L_\lambda = L_0 + \lambda \pi_s(q) \otimes \varphi_{AW}(\alpha) - \lambda \pi_s^\sharp(q) \otimes \varphi_{AW}^\sharp(\alpha), \tag{6.4}$$

where $\varphi_{AW}^\sharp$ denotes the field operator of the dual Araki-Woods representation $\pi_b^\sharp$ i.e.,
$$\pi_b^\sharp(\pi_{AW}(W(f))) = \exp\left(-i\varphi_{AW}^\sharp(f)\right).$$

For calculational purposes, let us develop a more explicit formula. Using the tensor product realizations $\mathcal{H}_s = \mathfrak{H}_s \otimes \mathfrak{H}_s$ and $\mathcal{H}_b = \mathfrak{H}_b \otimes \mathfrak{H}_b$, we can write

$$L_\lambda = L_s \otimes I + I \otimes L_b + \lambda (q \otimes I) \otimes \varphi_{AW}(\alpha) - \lambda (I \otimes q) \otimes \varphi^\sharp_{AW}(\alpha)$$

with

$$L_s \equiv h_s \otimes I - I \otimes h_s,$$
$$L_b \equiv h_b \otimes I - I \otimes h_b,$$

and

$$\varphi_{AW}(\alpha) \equiv \varphi((1+\rho)^{1/2}\alpha) \otimes I + I \otimes \varphi(\rho^{1/2}\bar\alpha),$$
$$\varphi^\sharp_{AW}(\alpha) \equiv \varphi(\rho^{1/2}\alpha) \otimes I + I \otimes \varphi((1+\rho)^{1/2}\bar\alpha).$$

We summarize the above discussion in

**Theorem 6.1.** *Let $(\mathcal{H}, \pi, \Omega)$ be the cyclic representation of the non-interacting spin-boson system $(\mathfrak{M}, \mathcal{S}_0^\beta, \tau_0)$ at inverse temperature $\beta < \infty$. For any $\lambda \in \mathbf{R}$ there exists a cyclic and separating vector $\Psi_\lambda^\beta \in \mathcal{H}$ such that*

$$\mathcal{S}_\lambda^\beta(A) \equiv \left(\Psi_\lambda^\beta, \pi(A)\Psi_\lambda^\beta\right),$$

*is the unique $(\tau_\lambda, \beta)$-KMS state of $\mathfrak{M}$. Furthermore the Liouvillean of the interacting spin-boson system is given by Equation (6.4).*

The Theorem 4.2 reformulates the problem of return to equilibrium as a spectral problem for the operator $L_\lambda$. Note that $L_0$ has the following spectrum:

$$\sigma_{ac}(L_0) = \mathbf{R},$$
$$\sigma_{sc}(L_0) = \emptyset,$$
$$\sigma_{pp}(L_0) = \{-2, 0, 2\}.$$

Clearly, $\pm 2$ are simple eigenvalues with eigenvectors $\psi_\pm \otimes \psi_\mp \otimes \Psi_b$, while $0$ is twofold degenerate eigenvalue with eigenvectors $\psi_\pm \otimes \psi_\pm \otimes \Psi_b$. When the interaction term is "switched on", one naturally expects all this eigenvalues to "turn into resonances", except for $0$ which must remain a simple eigenvalue. The precise way in which the degenerate eigenvalue $0$ splits into resonance and eigenvalue is the content of Einstein's A-B law.

The method developed in [JP1] for the analysis of the spectrum of the operator $H_\lambda$ immediately applies to $L_\lambda$. The reason is the following: The fundamental tool in [JP1] is the representation of $\mathcal{H}_b$ as the Fock space over $L^2(\mathbf{R} \times \mathrm{S}^2, ds\, d\sigma)$. In this representation, the interaction term in $H_\lambda$ is given by ([JP1], Theorem 3.1 and Equation 3.6)

$$H_I \equiv q \otimes \phi(\alpha_\beta),$$

where $\phi$ is the Segal field operator of the corresponding Fock representation of CCR, and

$$\alpha_\beta \equiv \left(\frac{s}{1 - \exp(-\beta s)}\right)^{1/2} \widetilde{\alpha} \in H^2(\delta, L^2(\mathrm{S}^2)),$$

as a consequence of Hypothesis (H2). A simple calculation shows that the corresponding term in $L_\lambda$ is

$$L_I \equiv (q \otimes I) \otimes \phi(\alpha_\beta) - (I \otimes q) \otimes \phi(\exp(-\beta s/2)\alpha_\beta),$$

which clearly enjoys similar analyticity properties.

Recall that $\Gamma^\beta_\pm$ and $\Pi^\beta_\pm$ were defined in Theorem 3.5. We further set

$$\Gamma^\beta \equiv \Gamma^\beta_- + \Gamma^\beta_+,$$
$$\Pi^\beta \equiv \Pi^\beta_+ - \Pi^\beta_-.$$

Then a simple adaptation of Theorem 2.2 and Proposition 4.7 in [JP1] gives

**Theorem 6.2.** *Suppose that Hypotheses (H1)-(H2) are satisfied. Then there exists a dense subspace $\mathcal{E} \subset \mathcal{H}$ and, for each $\eta \in ]0, \delta[$, a constant $\Lambda(\eta) > 0$ such that for $\lambda \in ]-\Lambda(\eta), \Lambda(\eta)[$ and $\Phi, \Psi \in \mathcal{E}$, the functions*

$$z \mapsto (\Phi, (L_\lambda - z)^{-1}\Psi), \tag{6.5}$$

*have a meromorphic continuation from the upper half plane onto the region*

$$\mathcal{O} \equiv \{z : \mathrm{Im}(z) > -\eta\}.$$

*The poles of matrix elements (6.5) in $\mathcal{O}$ are independent of $\Phi$ and $\Psi$. They are identical to the eigenvalues of a **quasi-energy** operator $\Sigma_\lambda$ on $\mathcal{H}_s$. This operator is analytic for $|\lambda| < \Lambda(\eta)$, with a power expansion of the form*

$$\Sigma_\lambda = H_s + \sum_{n=1}^\infty \lambda^{2n} \Sigma^{(2n)}.$$

*The matrix $\Sigma^{(2)}$ can be explicitly computed and, denoting by $P_E$ the eigenprojections of $H_s$, we have*

$$P_{\pm 2} \Sigma^{(2)} P_{\pm 2} = \left(\pm \Pi^\beta - i\Gamma^\beta\right), \tag{6.6}$$

*for the simple eigenvalues, and*

$$P_0 \Sigma^{(2)} P_0 = \begin{pmatrix} -2i\Gamma_-^\beta & 2i\Gamma_-^\beta e^{-\beta} \\ 2i\Gamma_+^\beta e^\beta & -2i\Gamma_+^\beta \end{pmatrix}, \tag{6.7}$$

*for the degenerate one. Note that the eigenvalues of the matrix (6.7) are $0$ and $-4i\Gamma^\beta$.*

An immediate consequence of the above result and of Proposition 4.1 in [CFKS] is that there is a constant $\Lambda(\beta) > 0$ such that for $0 < |\lambda| < \Lambda(\beta)$ the spectrum of $L_\lambda$ is purely absolutely continuous, except for the simple eigenvalue $0$. The proof of Theorem 2.3 is completed by invoking Theorem 4.2 and Corollary 4.3.

∎

**Remark 1.** We note that the matrix $\Sigma^{(2)}$ is intimately related to the generator of the Markovian dynamics that arises in the van Hove limit $\lambda \to 0$, $t = \lambda^{-2}\tau$. This generator is usually derived from Pauli's master equation (see for example [D1], [D2], [D3] or [M]). It turns out that, in the representation we work in, $-i\Sigma^{(2)}$ is *identical* to this generator. The relation between Pauli's master equation and the method developed in this paper will be discussed in more detail in [JP2].

We now sketch the proof of Theorem 3.4. We will only consider the limit $t \uparrow +\infty$, a similar argument can be used for $t \downarrow -\infty$. We invoke the dynamical consequence of Theorem 6.2 i.e., Theorem 2.5 in [JP1].

**Theorem 6.3.** *Suppose that Hypotheses (H1)-(H2) are satisfied. Then there exists a dense subspace $\mathcal{E} \subset \mathcal{H}$ and, for each $\eta \in ]0, \delta[$, a constant $\Lambda(\eta) > 0$ with the following property: For $|\lambda| < \eta$ there are two maps $W_\lambda^\pm : \mathcal{E} \to \mathcal{H}_s$ such that for any $\Phi, \Psi \in \mathcal{E}$, one has $(W_\lambda^- \Phi, W_\lambda^+ \Psi) = (\Phi, \Psi)$ and*

$$\left(\Phi, \exp(-itL_\lambda)\Psi\right) = \left(W_\lambda^- \Phi, \exp(-i\Sigma_\lambda t) W_\lambda^+ \Psi\right) + O(\exp(-\eta t)),$$

*as $t \to +\infty$.*

If $U(\theta)$ denote the group of translations introduced in Section 4 of [JP1], then $\mathcal{E}$ can be chosen to be the set of vectors which are analytic for $U(\theta)$ in the strip $\mathfrak{S}(\delta)$.

We define the set of states $\mathcal{N}_0$ and the algebra $\mathfrak{M}_0$ as follows. Let

$$\mathcal{A}_0 \equiv \{X \otimes \pi_{AW}(W(f)) : \tilde{f} \in H^2(\delta, L^2(S^2))\}, \tag{6.8}$$

denote by $\widetilde{\mathcal{N}}_0$ the set of vector states associated with vectors in $\pi^\sharp(\mathcal{A}_0)\Psi_\lambda^\beta$ and set $\widetilde{\mathfrak{M}}_0 = \pi(\mathcal{A}_0)$. We define $\mathcal{N}_0$ as the set of finite convex linear combinations of states in $\widetilde{\mathcal{N}}_0$ and $\mathfrak{M}_0$ as the linear span of $\widetilde{\mathfrak{M}}_0$. Clearly $\mathcal{N}_0$ and $\mathfrak{M}_0$ enjoy the properties stated in Theorem 3.4. Note also that it is sufficient to prove (3.4) for $\mathcal{S} \in \widetilde{\mathcal{N}}_0$ and $A \in \widetilde{\mathfrak{M}}_0$. Let $\gamma(\lambda)$ be the negative imaginary part of the complex eigenvalue of $\Sigma_\lambda$ closest to the real axis. Then it clearly satisfies the properties stated in the theorem. By Equation (4.4) and Theorem 6.3 the proof of Theorem 3.4 reduces to showing that the vectors $\pi(A)\Psi_\lambda^\beta$ and $\pi^\sharp(A)\Psi_\lambda^\beta$ belong to the set $\mathcal{E}$ for $A \in \mathcal{A}_0$. In the representation of $\mathcal{H}_b$ as the Fock space over $L^2(\mathbf{R} \times \mathbf{S}^2, ds\, d\sigma)$ used in [JP1], we have that

$$\pi(X \otimes \pi_{AW}(W(f))) = \pi_s(X) \otimes W\left(\frac{s}{1-\exp(-\beta s)}\tilde{f}\right),$$
$$\pi^\sharp(X \otimes \pi_{AW}(W(f))) = \pi_s^\sharp(X) \otimes W\left(\frac{s}{\exp(\beta s)-1}\tilde{f}\right).$$

Therefore, from our assumption on $\tilde{f}$ in Definition (6.8), Theorem 3.4 will follow from $\Psi_\lambda^\beta \in \mathcal{E}$. Using the notation of [JP1], this can be established as follows. Let $L_\lambda(-i\theta)$ be the deformed Liouvillean defined as in Section 4 of [JP1]. By Theorem 4.6 in [JP1], there is a constant $\Lambda > 0$, so that for $\frac{|\lambda|}{\Lambda} < \theta < \delta$ and $\Phi, \Phi' \in \mathcal{E}$, one has

$$(\Phi, \Psi_\lambda^\beta)(\Psi_\lambda^\beta, \Phi') = (U(i\theta)\Phi, Q_\lambda(-i\theta)U(-i\theta)\Phi'), \tag{6.9}$$

where $Q_\lambda(-i\theta)$ is the spectral projection of the deformed Liouvillean $L_\lambda(-i\theta)$ corresponding to the eigenvalue zero. Let now $\Phi = U(-i\theta)\Phi_0$ and $\Phi' = U(i\theta)\Phi_0$ with $\Phi_0$ an analytic vector for $U(\theta)$ in the strip $\mathfrak{S}(2\delta)$. Then $\Phi, \Phi' \in \mathcal{E}$, and from Equation (6.9) we conclude that

$$\left|(U(-i\theta)\Phi_0, \Psi_\lambda^\beta)\right| \left|(\Psi_\lambda^\beta, U(i\theta)\Phi_0)\right| = |(\Phi_0, Q_\lambda(-i\theta)\Phi_0)| \leq C_{\theta,\lambda} \|\Phi_0\|^2.$$

Thus for $\frac{|\lambda|}{\Lambda} < \theta < \delta$ we have

$$\Psi_\lambda^\beta \in \mathrm{D}(U(-i\theta)) \cap \mathrm{D}(U(i\theta)),$$

and therefore $U(\theta)\Psi_\lambda^\beta$ is analytic in $\mathfrak{S}(\delta)$. This completes the proof of Theorem 3.4.